# Learning-based super interpolation and extrapolation for speckled image reconstruction

HUANHAO LI[#1,2], ZHIPENG YU[#1,2], YUNQI LUO[#3], SHENGFU CHENG[1,2], LIHONG V. WANG[*4], YUANJIN ZHENG[*3], AND PUXIANG LAI[*1,2]

[1] *Department of Biomedical Engineering, Hong Kong Polytechnic University, Hong Kong SAR, China*
[2] *Hong Kong Polytechnic University Shenzhen Research Institute, Shenzhen, Guangdong, China*
[3] *School of Electrical and Electronics Engineering, Nanyang Technological University, Singapore, 639798*
[4] *Caltech Optical Imaging Laboratory, Andrew and Peggy Cherng Department of Medical Engineering, California Institute of Technology, Pasadena, California 91125, USA*
[#]*These authors contributed equally to the work*
[*]*Corresponding authors:* LVW@caltech.edu, yjzheng@ntu.edu.sg, *and* puxiang.lai@polyu.edu.hk

**Information retrieval from visually random optical speckle patterns is desired in many scenarios yet considered challenging. It requires accurate understanding or mapping of the multiple scattering process, or reliable capability to reverse or compensate for the scattering-induced phase distortions. In whatever situation, effective resolving and digitization of speckle patterns are necessary. Very often, a large field of view (FOV) is preferred to adequately encompass the encoded object information, which, however, inevitably leads to poorly sampled individual speckle grains with a limited-size optical sensor. The sampling insufficiency may cause the loss of correlation among speckle grains and hence the encoded object information, impeding successful object reconstruction from speckle patterns. Therefore, interpolation to poorly sampled, especially sub-Nyquist sampled, speckles is required to recover the lost information. For sub-Nyquist sampled speckles, we propose an InterNet, a deep learning-based network, to effectively interpolate them to well-resolved speckle patterns, which is impossible with classic interpolation methods. It shows that this learning-based interpolation approach provides a robust and promising framework to recover in high fidelity the lost object information and comprehensive speckle morphology from speckles that are poorly sampled (~14 times below the Nyquist criterion). Furthermore, the trade-off between the FOV and resolution of speckles can be favorably overcome as the framework equivalently improves the resolution by up to 32 times under the same FOV. Therefore, the learning network provides a new perspective on understanding the nature beneath speckles (e.g. how information is encoded in speckles) and a promising platform for efficient processing or deciphering of massive scattered optical signals, making it possible to see big and see clearly simultaneously in complex scenarios.**

http://dx.doi.org/10.1364/optica.99.099999

## 1. Introduction

Light experiences strong scattering in biological tissue, and the interference of multiply scattered photons traveling along different paths leads to the formation of speckles, if the coherence length of light is sufficiently long [1]. Information carried by light is therefore scrambled yet deterministically encoded in these speckles, [2] which are visually observed as spatially isolated bright spots, as long as the observation is completed within the speckle correlation time. Although challenging, it is feasible to analyze and/or retrieve the scrambled information through methods such as speckle correlation imaging [3-5], iterative wavefront shaping [6, 7], transmission matrix inversion [8-10], and optical phase conjugation [11-16]. For example, by tilting the incident beam within the angular range of the memory effect, the varying speckle patterns are highly correlated, which can be used to enlarge the field of view (FOV) and improve the resolution via a Gerchberg-Saxton-type algorithm [5, 17]. A hidden object can be also recovered from the autocorrelation of one camera image, whose optical intensity profile is randomized by the scattering medium, when a Fienup-type algorithm is applied [3]. Another approach is based on the measured transmission matrix (TM) in weak scattering regime: optically, the singular vectors obtained through singular value decomposition (SVD) are linked to the distribution of the optical scatters [18], and photoacoustically, distribution of optical absorbers can be also demodulated from the TM and hence improve the quality of photoacoustic imaging [19].

Deep learning is another powerful tool for speckled information retrieval or reconstruction and has more recently seen rapidly development and huge potentials[20]. In deep learning, stacked deep neural network (DNN), a layered structure with linear/nonlinear computational units optimized by gradient descent methods [21], can automatically learn how to extract different levels of feature representation from the raw data, so that a complex function can be learned [22]. Due to its featured training performance, an end-to-end solution, i.e. raw data input to expected output, can be directly provided by deep learning without feature extractor or excessive preprocessing. The idea was first demonstrated for visualizing an object from its diffracted patterns, where a DNN was trained to learn and decipher the diffraction intensity profile [23]. Optical scattering was later involved and a phase object behind a ground glass was recovered from the intensity profile of speckle patterns [24]. Speckle patterns arising from multimode fibers can also be learned to reconstruct phase objects positioned at the other end [25, 26]. These 'one-to-one' achievements confirm the ability of DNNs to learn the complex scattering transformation of the medium on investigation. Furthermore, deep learning can also generalize the transformation to multiple media ('all-to-one') through feeding speckle patterns from a single object but randomized by different disordered media into a single trained DNN [27].

These successful image reconstructions based on deep learning can be attributed to a large number of speckle grains included in the recorded images. And more detected speckles, or a larger FOV, can better encompass information encoded in the speckle pattern [28]. Effective detection for speckles is however limited by the trade-off between resolution and FOV of speckle patterns recorded with digital cameras: more camera pixels to resolve individual speckle grains may lead to insufficient FOV; less pixels for each grain can improve the acquisition speed and reduce the data size, but the essential object information may be lost if the correlation among speckle grains is not sufficiently sampled. Another possible measure is to sample more speckle grains to increase the FOV (increase the effective information), which, however, requires more camera pixels (further slows down the acquisition) or resolves individual speckle grains in a sub-Nyquist domain due to limited camera pixels. A recent work from Shen et al. finds out that sub-Nyquist sampled speckles boost the performance of optical focusing via wavefront shaping [29], but how different spatially sampled speckles affect the computational image reconstruction has been rarely explored. This spurs our hypotheses: Can spatially down-sampled speckle patterns be morphologically recovered to well-resolved ones? More importantly, can the recovered speckle patterns retrace the lost information due to insufficient sampling?

Recent progresses in the ballistic regime [30, 31] have shown that DNNs can transform diffraction-limited images into super-resolution ones, such as from confocal microscopy to stimulated emission depletion (STED) microscope and from total internal reflection fluorescence (TIRF) microscopy to structured illumination microscopy (SIM). Therefore, the feasibility of our hypothesis seems conceptually foreseeable with deep learning. But challenges exist for down-sampled speckles. In the ballistic regime, the point spread function (PSF) is clearly defined, and the mapping between the object and the obtained image is localized (or short-range correlated) and 'one-to-one'. Thus, the obtained image directly reflects the morphology of the object, albeit with some degrees of localized distortions due to the medium and/or the system. For speckled images, however, the PSF is distorted by scattering; one point in the sample (object) plane contributes to a many of points in the speckle recording (image) plane [9, 32], i.e., 'all-to-one'. Therefore, information encoded between the object and the corresponding speckle pattern is delocalized (or long-range correlated) [2],[33], which is spatially distributed everywhere within the FOV in a form of grainy morphology rather than a blurred object. As a result, speckle patterns that appear visually akin, e.g. of high correlation coefficient, are not necessarily linked with the same object; the existence of an object needs further validation. To the best of our knowledge, so far, few studies, if any, have explored whether or not down-sampled delocalized information can be interpolated or extrapolated (for simplicity, both operations will be termed as interpolation in this work) and retrieved through deep learning.

In this work, we introduce an implementation of DNN, namely InterNet, to interpolate down-sampled speckle patterns to well-resolved speckle patterns and retrieve the lost information. The retrieval is validated by further feeding the interpolated speckles into another DNN, namely SpeckleNet, to accomplish generalized image reconstruction. Experimental results show our learning-based nonlinear interpolation can handle very sparsely sampled speckles (down to ~1/14 of the Nyquist criterion) and equivalently improve the resolution of speckles by a factor of up to 32 under the same FOV. As a result, not only the comprehensive morphology of the speckle patterns but also the lost delocalized information can be effectively recovered, allowing for robust object reconstruction. Notably, the well-trained InterNet generalizes and recovers the delocalized information encoded in the speckles by merely learning speckles sampled at different frequencies, with no object image involved in the training. Nevertheless, the InterNet can still learn how information is encoded in the speckles and lost due to spatial sampling, with well resolved speckles only as the target for training. Therefore, this study provides a new perspective to understand the nature beneath speckles and a powerful platform to process or decipher delocalized speckled signals.

## 2. Method

**Experimental setup.** The experimental apparatus is configured as in Fig. 1. Images extracted from MNIST database are used as the phase objects and are displayed on a phase modulation spatial light modulator (SLM, HOLOEYE PLUTO VIS056 1080p, German). A collimated continuous-wave coherent laser beam at $\lambda = 532$ nm (EXLSR-532-300-CDRH, Spectra Physics, USA) is expanded, so that the screen of the SLM is fully illuminated. The laser beam is modulated by the SLM with phase objects uploaded in advance. Since the handwritten digits in MNIST are 28-by-28 in terms of pixels, these images are up-sampled to 1024-by-1024 pixels and displayed on the SLM sequentially. The phase-only SLM converts 8-bit grayscale (0-255) to phase delay (0-$2\pi$ in radian). As the images from MNIST are also quantified by the 8-bit grayscale, they are rescaled from 0-255 to 0-127 to strengthen the modulation efficiency. After being modulated and reflected by the SLM, the laser beam is converged onto a diffuser (220-grid, DG10-220-MD, Thorlabs, USA) by an objective (RMS20X, Olympus, Japan). Totally 20,000 images from MNIST database are sequentially displayed on SLM. the corresponding scattering-induced speckle patterns are one by one captured by a CMOS camera (FL3-U3-32S2M-CS, PointGrey, Canada). The camera and the SLM are synchronized via a MATLAB program during data acquisition.

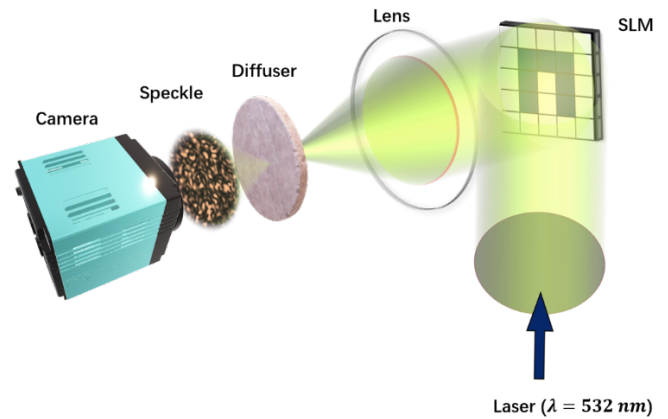

Fig. 1 Experimental setup for speckle patterns collection and object reconstruction: phase objects are displayed on the spatial light modulator (SLM), which is illuminated by an expanded continuous coherent laser beam ($\lambda = 532$ nm).

**Nyquist criterion and data preparation.** The recorded speckle patterns and the corresponding images displayed on the SLM are paired up as a dataset for DNN training, with speckle pattern as the input and the corresponding image as the output (target). In this study, 20,000 samples for each sampling factor are generated, 19,800 of which are used for DNN training while the rest for network validation. The sampling factor (F), to represent the pixel number occupied by one speckle grain on average, is defined by the area of full width at half maximum (FWHM) regarding the autocorrelation of speckle patterns [34]. The well-sampled speckle patterns with an initial sampling factor ($F_0$= 17 pixels) are captured by the camera and sampled by 256-by-256 pixels, as shown in Fig. 4a. The pixel pitch of the camera is 2.5 $\mu m$, thus the average diameter of speckle grain can be calculated: $\pi(D/2)^2 = 17 \times 2.5^2 \mu m^2 \Rightarrow D \approx 11.63\ \mu m$. Due to the Nyquist criterion [35], the pixel size is conventionally chosen as a half of the minimum spacing. Notably, since it is difficult to find out the smallest size of speckle grain which, however, is typically represented by the average value [34], we define the cut-off Nyquist sampling size of camera pixel is a half of the average diameter of speckle grains, i.e. $d_c = D/2 \approx 5.82\ \mu m$. For simplicity and ensuring differently sampled speckle patterns are evaluated under the same experiment conditions, the down-sampled speckle patterns are computationally obtained through pixel binning [29]. Thus, the sampling pitch (ds) can be equivalently defined as $2.5 \times n\ \mu m$ by grouping n-by-n neighboring camera pixels as a macropixel. As shown in Fig. 4a-h, the speckle patterns are sampled with ds =2.5, 5, 10, 20, 40, 80, 160, and 320 $\mu m$, respectively, so that the relative pitch with respect to $d_c$ are 0.43, 0.86, 1.72, 3.44, 6.87, 13.76, 27.49, and 54.98 (denoted as di, i=0, 1, ..., 7), correspondingly. Therefore, the first two sampling pitches meets the Nyquist criterion, but the others belong to sub-Nyquist sampling. Based on these eight selected sampling pitches, eight datasets, or called di-dataset (i = 0, 1, ..., 7), are generated, including eight training datasets (19,800 samples for each ds) and eight test datasets (200 samples for each $d_s$). Note that the speckle-digit pairs in the training and test sets have no overlap.

**DNN architecture and training.** DNN architectures based on DeeplabV3+ [36] are used in this study for speckle interpolation and image reconstruction by properly assigning the input and output for the neural networks, as shown in Fig. 2. The developed DNN for image reconstruction is named SpeckleNet and DNN for interpolation is denoted as InterNet. SpeckleNet is trained by one dataset ($d_0$ sampled speckle pattern as input and digit from MNIST as output) through minimizing a loss function of the negative Pearson's correlation coefficient (NPCC). For interpolation, Type 1 (InterNet-1) is the standard DeeplabV3+ but Type 2 (InterNet-2) is a modified DeeplabV3+ with densely connected layers to replace the Bilinear interpolation in the standard version. Both InterNets are individually trained by 5 datasets, in which training sets in di-dataset (i = 3,4,5,6,7) act as the input and corresponding samples in training sets in $d_0$-dataset as the output. As all InterNets aim to interpolate the down-sampled speckle patterns to $d_0$-sampled patterns, they are simply denoted as di-trained InterNet if di-dataset is used for training. For simplicity, the main text ignores the expression of 'di-trained' for all InterNets. Two loss functions are trialed to optimize the processing. InterNets trained with a loss function of a negative Pearson's correlation coefficient (NPCC) is denoted as InterNet(cc), and InterNets trained with a combination loss function of NPCC and mean square error (MSE), denoted as comloss in the main text, is called InterNet(com). Assuming y is the target of DNN and $\hat{y}$ is the output from DNN with average operation $\langle\ \rangle$, variance $\sigma$ and absolute operation $\|\ \|$, both loss functions can be formulated as:

$$NPCC(y,\hat{y}) = -\frac{\langle(y-\langle y\rangle)(\hat{y}-\langle\hat{y}\rangle)\rangle}{\sigma_y \sigma_{\hat{y}}} \quad (1)$$

$$MSE(y,\hat{y}) = \langle\|y-\hat{y}\|^2\rangle \quad (2)$$

$$comloss(y,\hat{y}) = -\frac{\langle(y-\langle y\rangle)(\hat{y}-\langle\hat{y}\rangle)\rangle}{\sigma_y \sigma_{\hat{y}}} + \langle\|y-\hat{y}\|^2\rangle \quad (3)$$

It should be clarified that although the same DNN structure is implemented for both interpolation and imaging reconstruction, the training for two tasks are independently conducted, and the parameters in InterNets and SpeckleNets will not be affected by each other. Such isolated training provides the feasibility that InterNets can be trained for merely interpolation without knowing the object information. In practice, detected signal from the scattering medium is merely speckle pattern and the imaging target is unknown. Since the SpeckleNets need imaging target, the simultaneous training of them stops the InterNets from its practical applications. In training, all DNNs are optimized by 100 epochs with batch size of 16 and the learning rate decays from 0.05 following the cosine annealing strategy during the back-propagating gradient descent optimization. The training framework is Pytorch 0.4.0 with python 3.7, using CUDA for GPU acceleration. Computing unit is a Dell precision workstation with E5-1620v3, 56 Gb RAM, and a RTX2080 ti GPU.

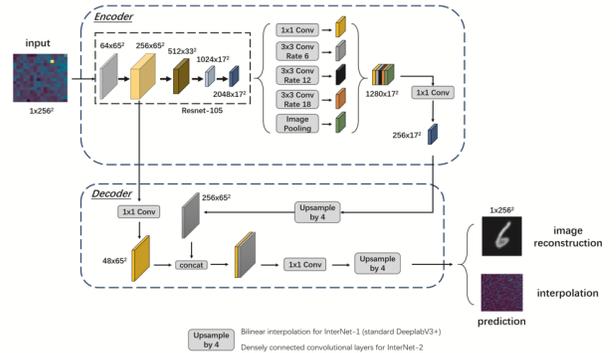

Fig. 2 Architectures of neural network based on DeeplabV3+ with a backbone of ResNet-105 [36]. InterNet-1 and SpeckleNet in this study are the standard version, whose upsampling operation in decoder is 4x Bilinear interpolation. InterNet-2 replaces the 4x Bilinear upsampling with a densely connected convolutional layers. For image reconstruction (SpeckleNet), the output of the network is the digit from MNIST and for interpolation (InterNet) the output is $d_0$-sampled speckle pattern. Inputs for both tasks are speckle patterns.

**Classic interpolation.** Three classic interpolation methods are selected as comparison in this study. The nearest neighbor algorithm (denoted as Nearest in the main text) is based on zero-order polynomial (constant). It inserts values at the interpolated points between existing points by value of the nearest data point and no new data is generated in the interpolated points. The Bilinear and Bicubic interpolation, however, generate new data in the interpolated points. The Bilinear interpolation fits a linear polynomial (first-order) between each pair of existing points and the Bicubic fits a cubic polynomial (third-order) between the existing points.

**Workflow.** The workflow of interpolation and image reconstruction is shown in Fig. 3. Speckle patterns sampled from $d_0$ to $d_2$ are directly fed into the SpeckleNet for image reconstruction. For speckle patterns sampled by other five sampling pitches ($d_3$ to $d_7$), classic and DNN based interpolations will be applied to the sub-Nyquist sampled speckle patterns (from poorly sampled speckles to $d_0$ sampled speckles). The interpolated patterns will be first investigated regarding the morphology. To validate whether the interpolation process regains the object information (e.g. Fig. 4q) as the exampled well-resolved speckle pattern (e.g. Fig. 4a), the interpolated patterns are also fed into the

SpeckleNet. Metrics, including the PCC and MSE between the interpolated and the target speckle patterns, are quantified for analysis. To be noted that, SpeckleNet here is only used to validate the existence of the information and the interpolation is not necessarily used for image reconstruction.

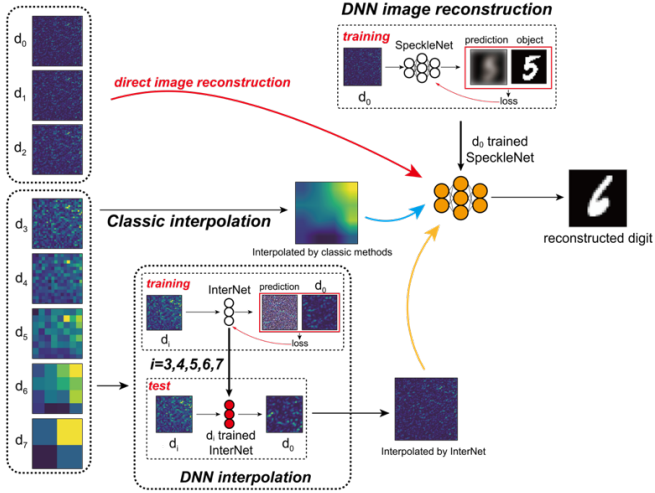

Fig. 3 Workflow diagram of learning-based interpolation and image reconstruction for speckle patterns with different sampling frequencies. SpeckleNet is only trained by $d_0$-sampled dataset (definition in Methods). $d_0$-$d_2$ sampled speckle patterns are directly fed into the trained SpeckleNet to achieve successful image reconstruction. InterNets, regardless of different loss functions, are individually trained by inputting different di- sampled speckle patterns (i = 3, 4, 5, 6, 7) and outputting the $d_0$- sampled speckle patterns. InterNet with input of di-sampled speckle patterns (i = 3, 4, 5, 6, 7) is denoted as di trained InterNet (simplified as InterNet in the main text). The $d_3$-$d_7$ sampled speckle patterns are preprocessed by InterNet for interpolation and then fed into SpeckleNet for image reconstruction. To be noted that, the red arrows do not mean concurrent input of the SpeckleNet.

## 3. Speckles sampled with different frequencies

Fully resolved speckle patterns are collected by camera at first (Fig. 4a) based on the experimental apparatus as described in "Experimental Setup" of "Methods". The down-sampled speckle patterns are obtained digitally by pixel-binning (see Nyquist criterion in Methods) as shown in Fig. 4b-h. The relative sampling pitch $d = d_i$ (i = 0, 1, ..., 7) indicates the relative span between two neighboring sampling points ($d = d_s/d_c$, $d_s$ is the reciprocal of the sampling frequency and $d_c$ is Nyquist criterion), and their values corresponding to Fig. 4a-h are calculated and shown in Fig. 4r, among which $d_0$ and $d_1$ are above the Nyquist criterion, and the critical/cut-off sampling pitch dc = 5.82 μm (d>1, see Nyquist criterion in Methods). To quantify the differentiation due to the varied sampling, mutual correlation ($C_m$) is defined: it is the average Pearson's Correlation Coefficients between every pair of speckle patterns among the datasets for each specific d (20,000 samples for each sampling pitch). As shown in Fig. 4r, $C_m$ shows an increasing tendency with the sampling pitch (d) and approaches to one as sparsely sampled speckle patterns diminish the nuances and become more and more alike with each other, even though the displayed values on the SLM are different. When sampling the speckle patterns around and above the Nyquist criterion ($d_0$-$d_2$), $C_m$ is below 0.7 and it tends to converge to ~0.67 for smaller sampling pitches. And such high remaining $C_m$ values should be attributed to the fact that the object (digit 6 from MNIST, Fig. 4q) only shows non-zero values around the central areas (white color, <20% of total pixels). The rest zero-value pixels are displayed as null phase delay on the SLM as the background, leading to a high-level residual correlation among speckle patterns. It shows clearly that with higher sampling pitch to sample the speckle patterns, the distinguished features and grainy representation of the captured speckle patterns are bleached.

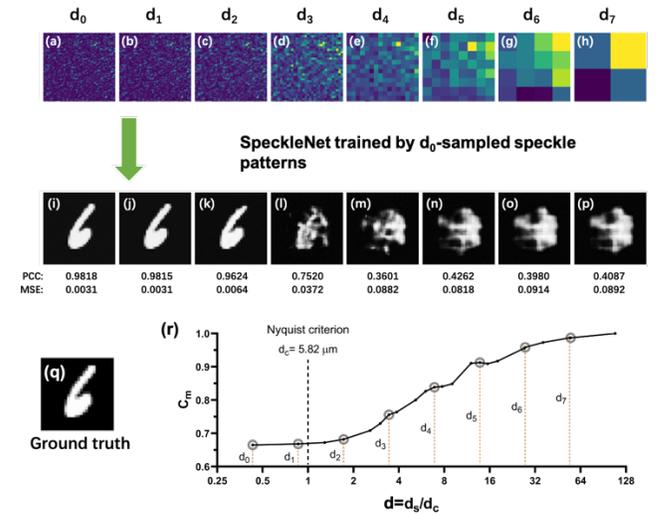

Fig. 4 Representative speckle patterns sampled with different pitches and the corresponding image reconstruction via deep learning. (a) Well-resolved (above Nyquist criterion) speckle pattern is originally captured a camera; (b)-(h) down-sampled speckle pattern from (a) by pixel binning; (i)-(p) are the reconstructed images predicted by feeding (a)-(h) into SpeckleNet that is trained by the $d_0$-sampled speckle pattern; (q) is the ground true target. PCC: Pearson's correlation coefficient between the output and (q); MSE: mean square error between the output and (q). (r) Mutual correlation ($C_m$) increases with the relative sampling pitch ( d ). $d_s$: the sampling pitch; $d_c$: the critical/cut-off sampling pitch given by the Nyquist criterion; $C_m$: the average of Pearson's Correlation Coefficients between every pair of speckle patterns among the datasets for each specific sampling pitch.

## 4. Information Validation

The aforementioned speckle patterns sampled at different pitches are then fed into a DeeplabV3+ [36] based DNN architecture for image reconstruction to validate whether or not the object information is included in the down sampled speckles. This DNN, named SpeckleNet, was trained with the originally captured speckle patterns ($d_0$-sampled) of the ground true target from MNIST (from speckles to object). More results about SpeckleNets trained by different d-sampled speckle patterns as well as comparison with the other widely used DNN architecture, U-net, can be found in Section 1 of Supplementary. Pearson's correlation coefficient (PCC) and mean square error (MSE) are the metrics to evaluate the statistical similarity and error between the prediction and the target: higher PCC and lower MSE indicate less differences and better reconstruction. Speckle patterns sampled at different pitches (Fig. 4a-h, 8 test sets, 200 samples for each pitch, and no overlap with the training sets as described in Method) are individually fed into the SpeckleNet; their corresponding outputs are the reconstructed digits (Fig. 4i-p).

As seen, the SpeckleNet successfully reconstructs the digit '6' (one example in test dataset) when the $d_0$-sampled speckle patterns are tested (Fig. 4i). For this group, the speckle pattern sampling pitch for the training set and the test set are the same. Such a result is consistent with earlier reported demonstrations [24, 26], although different DNN architectures are used. When there is certain sampling pitch inconsistence between the training and the test sets, the SpeckleNet shows some generalization capability: it is feasible to reconstruct digits that are visually the same (Fig. 4i-k) despite different sampling pitches and quantitatively from $d_0$ to $d_2$, the PCC reduces from 0.9818 to 0.9624 and the MSE increases from 0.0031 to 0.0064. Such effective generalization suggests that pixel-binning from $d_0$ to $d_2$ (16-fold compression) can actually be treated as a lossless compression. In Fig. 4c, with $d_2$-sampling, the fine features of speckle patterns are almost lost (sub-Nyquist sampling already) and one speckle grain is resolved by less than one image pixel on average. At this moment, information remained in the down-sampled speckle patterns mainly reflects the spatial distribution of speckle grains, i.e., the long-range correlation [37]. Similar levels of $C_m$ from $d_0$ to $d_2$-sampling indicate that the pixel-binning process does not break the long-range correlation but maintains the essential object information encoded in the speckle pattern. Therefore, these down-sampled speckle patterns (Fig. 4a-c) can be identified by the SpeckleNet. However, as the sampling pitch of tested samples further increases, the network generalization capability is significantly weakened. Only unidentifiable features can be reconstructed (Fig. 4l-p) with position consistent to that of the displayed digits (around the center). Quantitatively, larger mismatch induces worse metrics (lower PCC and higher MSE), compared with the ground true digit (Fig. 4q).

These results suggest the SpeckleNet, essentially a convolution neural network (CNN), seems to inherently possess interpolation ability, so that the trained network can have generalization for differently sampled speckle patterns. To confirm that, the architecture of DeeplabV3+ based SpeckleNet is shown in Fig. 2. As seen, down-sampling operation is included in the encoder part (ResNet-105) through two-dimension (2D) average pooling layers. Notably, the 2D-average pooling operation in DNN functions exactly the same as the pixel-binning, indicating that sub-Nyquist sampled information can be generated within the DNNs. However, due to the limited depth of DNNs, the poorly sampled information is processed by less layers. Adding more layers may overcome the drawback, but considerably increase the computational burden. On the other hand, between two down-sampling operations, the information within the encoder is transformed by the layers with learnable parameters, instead of a non-parametric pixel binning or 2D average pooling. Thus, the processed feature maps are highly related to the sampling information of the training data, $d_0$-sampled speckle pattern in this case. For sub-Nyquist sampled speckle patterns, the information nuance encoded by different objects is smoothed, causing significant increase of $C_m$ when $d > d_2$. In this case, the $d_0$ − trained SpeckleNet can no longer generalize image reconstruction, unless the smoothed or binned features in the speckle patterns can be recovered through interpolation.

In the following sections, the **SpeckleNet will act as the tool to validate** whether or not the object information is contained in the interpolated speckles.

## 5. Classic interpolation

Three classic interpolation methods, namely Bilinear, Bicubic, and Nearest, are selected to interpolate the sub-Nyquist sampled speckle patterns. For $d_3$-sampled speckle patterns, grainy representations can still be observed after interpolation (Fig. 5a); with higher sampling pitches, the grainy morphology is further degraded after interpolation, left with only one obscure grain on the top right corner (Fig. 5b-d). Such fading is highly correlated with the degradation of PCC. For example, with Bilinear interpolation, more fluctuations are observed for speckle patterns with lower sampling pitches (Column II). The trend is quite similar with the other two interpolation methods, but higher orders of interpolation lead to more grainy morphological features. Comparably, the MSE does not show a clear dependence on the sampling pitch, and it is identical before and after the classic interpolations no matter the speckle patterns are visually akin (e.g., Columns I and VI, Columns II and IV) or not (e.g., Columns IV and V). Thus, same values of MSE do not ensure consistent speckle morphology (e.g. MSE = 0.0086 for all speckle patterns in Fig. 5c-d); PCC seems to characterize the morphology of speckle patterns or local structures [38] better than the MSE. But neither of them can exclusively define a unique speckle pattern, which will be further discussed later.

The interpolated speckle patterns (Columns II, IV, and VI in Fig. 5) are all fed into the SpeckleNet. The realizations are similar to those in Fig. 4l-p, indicating that these interpolated patterns can't be recognized by the SpeckleNet to reveal the hidden object. Therefore, for sub-Nyquist sampled speckle patterns, the classic interpolations cannot retrieve or add object information to the speckle patterns for image reconstruction, regardless of the sampling pitch and interpolation method. To double confirm this, speckle patterns encoded with digits '2', '0', '1', '9' are also tested and illustrated in Fig. 7b-d with consistent performance.

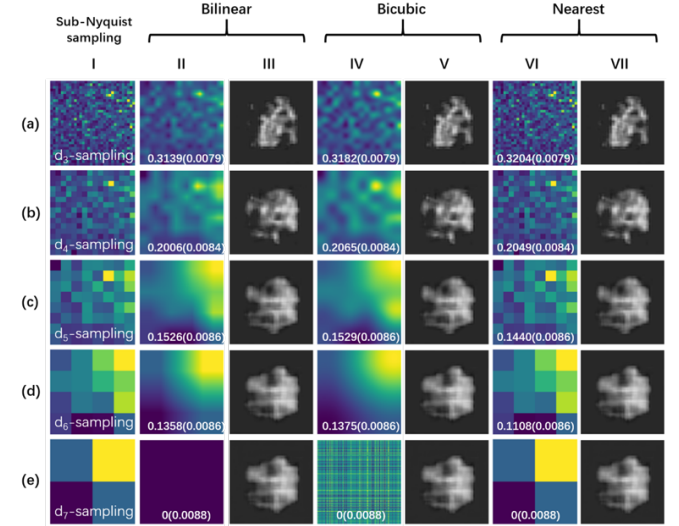

Fig. 5 Interpolation based on classic methods and the corresponding learning-based imaging reconstruction. Sub-Nyquist sampled speckle patterns, from $d_3$ to $d_7$, are aligned in I(a-e). Column II, IV and VI are the interpolated patterns based on Bilinear, Bicubic and Nearest Interpolation, respectively, and the inset quantities are formatted as PCC (MSE). Column III, V and VII are the reconstructed images predicted by feeding Column II, IV and VI speckles into SpeckleNet. PCC: (negative) Pearson's correlation coefficient between the interpolated speckle pattern and the reference fully-sampled speckle pattern (Fig. 4a); MSE: mean square error between the interpolated output and the reference (Fig. 4a).

## 6. DNN-based interpolation

A DNN, called InterNet, is investigated for interpolation. Each down-sampled speckle pattern (as network input) with sampling pitch from $d_3$ to $d_7$ is paired up with the corresponding $d_0$-sampled speckle pattern (as network output) to train the InterNets. The network is optimized through minimizing two different loss functions: 1) a combination loss of NPCC and MSE (comloss) and 2) NPCC only. Note that the InterNet is independently trained so that the reconstructed patterns from the SpeckleNet won't be used to update the parameters of the InterNet. Four InterNets are trained and denoted as InterNet(com) and InterNet(cc), respectively, depending on whether comloss or NPCC is used as the loss function. On the other hand, the InterNet-1 is based on the standard DeeplabV3+ and the InterNet-2 is a modified DeeplabV3+ with densely connected layers to replace the Bilinear interpolation in the standard DeeplabV3+. For more details about DNN training, readers can refer to Methods.

The InterNet training-based interpolation results are shown in Fig. 6 II, IV, VI & VIII. As seen, all four InterNets are able to recover the grainy features of the five sub-Nyquist sampled speckle patterns; the randomly scattered bright spots can be clearly identified in all the realizations, which outperforms the classic interpolation predictions (Fig. 5 II, IV & VI) apparently. The morphological features of these interpolated patterns are similar but the tiny distinctions can be statistically described by the PCC between the interpolation and target (Fig. 4a). For more detailed morphological observations, zoom-in interpolated patterns and intensity profiles, please refer to Section 2 of Supplementary. It shows that with the same interpolation method, the PCC decreases slightly yet monotonically with the original sampling pitch. The trend is similar to that observed with the classic interpolation methods (Fig. 5), but the PCCs with DNN-interpolation are significantly higher. For example, Fig. 6Ie speckle pattern is sampled at a frequency about 55 times less than the Nyquist criterion. Using the InterNets, the interpolated patterns have high PCCs with fully-resolved pattern (Fig. 4a), being 0.7297(Fig. 6 IIe), 0.7328 (Fig. 6 IVe), 0.8047 (Fig. 6 VIe) and 0.8028 (Fig. 6 VIIIe). With classic interpolation methods, the results are all 0 (Fig. 5IIe, IVe, and VIe).

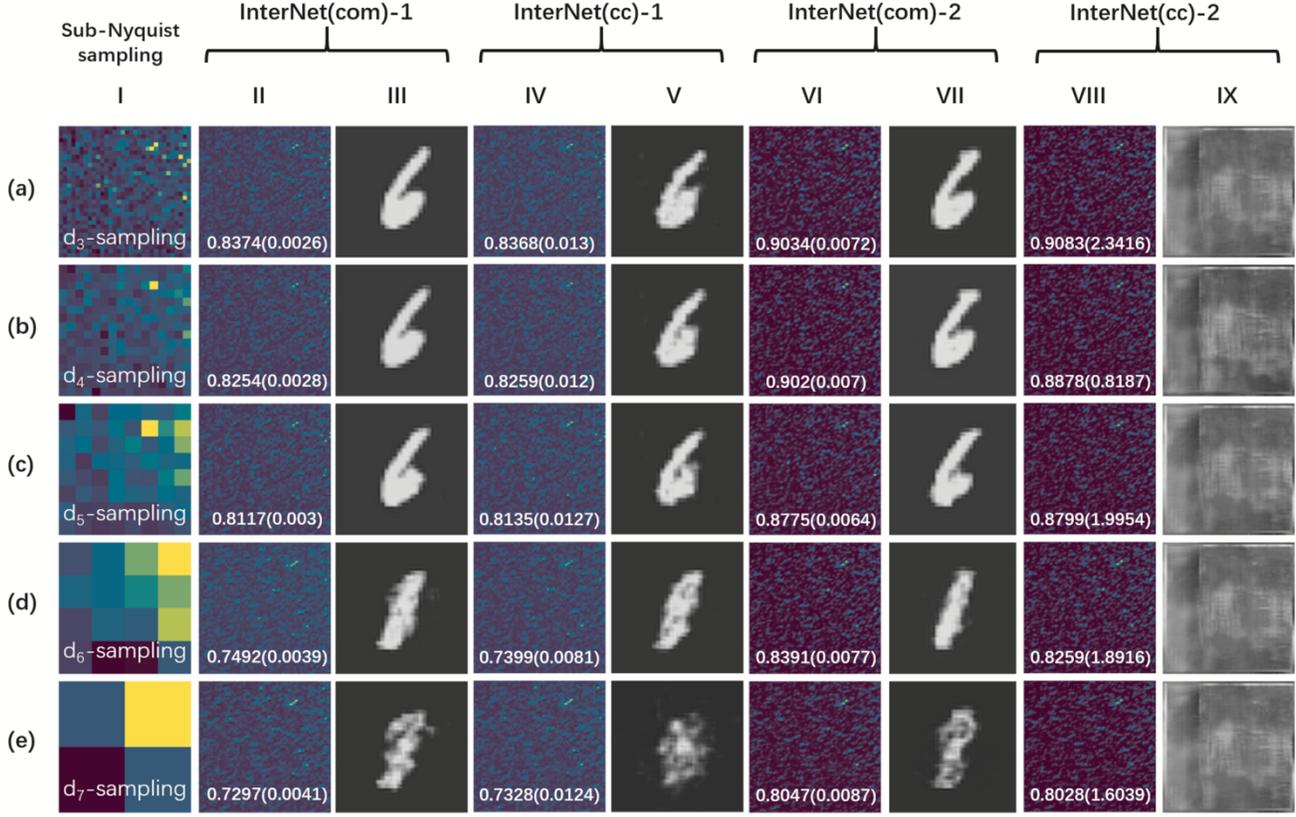

Fig. 6 Interpolation by InterNets and its corresponding imaging reconstruction. Sub-Nyquist sampled speckle patterns, $d_3$-$d_7$, are aligned in I(a)-I(e). Columns II, IV, VI and VIII are the InterNet-interpolated speckle patterns, with inset quantities formatted as PCC (MSE); Columns III, V, VII and IX are the reconstructed images by feeding Columns II, IV, VI and VIII into the SpeckleNet. InterNet(com) is InterNet trained by a combination loss (NPCC+MSE) and InterNet(cc) is InterNet trained by NPCC only. PCC: Pearson's correlation coefficient between the interpolated output and target (Fig. 4a); MSE: mean square error between the interpolated output and target (Fig. 4a).

Next, the InterNet-interpolated sub-Nyquist sampled patterns are fed as the inputs to the SpeckleNet to predict the encoded object information (digit '6' displayed on the SLM). Results are shown in Fig. 6 III, V, VII & IX, where the object is successfully reconstructed in some realizations. For speckle patterns with original sampling pitch less than $d_6$, interpolation by both InterNet-1s allows for effective image reconstruction by the SpeckleNet (Fig. 6 IIIa-c and Va-c). Similar performances are also supported by InterNet(com)-2 (Fig. 6VIIa-c), the other DNN with minor modifications.

These achievements demonstrate that a well-trained InterNet can interpolate sparsely sampled speckle patterns sufficiently well to retrace optical information randomized by strong scattering. The lowest sampling pitch supported in this study is more than one order (Fig. 4r, $d_5/d_c \sim 13.75$) below the Nyquist criterion, and the corresponding speckle patterns are interpolated by a factor of 32 (Fig. 4r and Fig. 6Ic, $d_5/d_0 = 32$, resolution is equivalently increased by 32 times). Imagine that for a two-dimensional image, data storage and transfer consumption can be saved by $32^2 = 1{,}024$ times with such a compression method through pixel binning! On the other hand, it provides feasibility to overcome the comprise between FOV and resolution. Notably, object information is successfully retraced (Fig. 6c III, V & VII) from the speckles poorly sampled by a large sampling pitch $d_5 \sim 13.75 d_c$. As seen in Fig. 6Ic, the grainy morphology of $d_5$ sampled speckles have been entirely smoothed out because signal at each pixel is an integration of several speckle grains, equivalently $32^2 = 1{,}024$ pixels in $d_0$ sampled speckles, and the object information is gone. And the camera detection in this study is 16 bit and the 8x8 FOV of $d_5$ sampling therefore has $[(2^{16})^{1024}]^{8\times 8} = 2^{2^{20}}$ possibilities for interpolating to the 256x256 $d_0$ sampling. These two seemingly impossible missions are accomplished by the proposed InterNet with supervised learning. Guided by the training set, the InterNets are equipped with the ability to resolve and up sample the $d_5$ sampled patterns, equivalently reversing the down sampling operation (i.e., pixel binning).

Such "super" interpolation capability of the first three InterNets, as shown in Fig. 6, fades for larger sampling pitches ($d > d_5$); the digit '6' can no longer be recovered but left with non-zero values around the center. The long-range correlation of speckle patterns is probably undermined when the speckle pattern sampling pitch is too high. Therefore, although the short-range correlation (the grainy morphology) can be retrieved by the InterNets (Fig. 6 II, IV, VI & VIII), the object recovery is more challenging if too much information has been filtered by pixel binning during the down-sampling process. For example, only of 49.5% and 37.5% of all samples in the test set can be reconstructed by InterNet(com)-1 and InterNet(com)-2, respectively, as shown in Section 4 of Supplementary. As for $d_5$-sampled speckle patterns, 93% in the test set can be effectively interpolated by InterNet(com)-1 and 89.5% by InterNet(com)-2 (Section 4 of Supplementary). Some of the realizations are shown in Fig. 7e-g, $d_5$-sampled speckle patterns encoded with digits '2', '0', '1', '9' can also be effectively up-sampled to $d_0$-pitched patterns as those shown in Fig. 4a by InterNet(com)-1 (Fig. 7e), InterNet(cc)-1 (Fig. 7f) and InterNet(com)-2 (Fig. 7g). Interpolation based on InterNet(cc)-2, however, fails to recover any of the sub-Nyquist sampled object information, such as digit '6' in Fig. 6IX and digits '2', '0', '1', '9' (Fig. 7h), which will be further discussed in the next section.

Nevertheless, the proposed learning-based interpolation approach provides a robust and promising solution to retrieve object information from down-sampled speckle patterns. Speckle patterns are featured by randomly distributed bright spots of varied sizes [1]. Such inhomogeneity prevents classic interpolation methods from recovering the morphology of speckle grains due to its global order consistency (see Methods), so that the object information filtered by the pixel binning process cannot be retrieved or decoded with whatever image reconstruction method. The inherent nonlinear activations in the DNNs, however, provide possibility to overcome this weakness. The DNNs allow interpolation with various orders of polynomial by adaptively inserting the values to form the morphological feature of speckle grains. Within one speckle grain, the information (including both intensity and phase) is highly correlated, which is termed short-range correlation [37]. Hence, learning-based interpolated patterns are able to consider or, more appropriately, learn the morphology of speckle grains beyond merely producing uncorrelated random signals. Such distinguished capability should be attributed to the supervised configuration of the deep learning method, which is trained by a dataset of 19,800 samples. With supervised learning, target in the training dataset specifically guide the parameters in the DNNs to update oriented to the direction. By minimizing the discrepancy between the network output and the target through back-propagation gradient descent, the information from the target flows back across each layer and updates their parameters. That's why the same neural network utilized in this study (Fig. 2) can be used for different tasks, if the targets in the training data set are different.

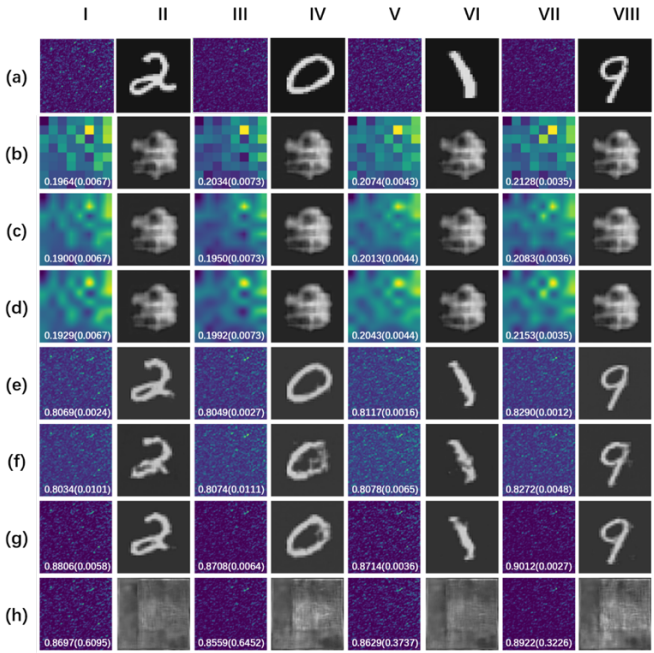

Fig. 7 Additional realizations of DNN-based interpolation and image reconstruction for sub-Nyquist sampled speckle patterns. Speckle patterns containing the information of digits '2', '0', '1', '9' are aligned in Columns I, III, V, and VII; the corresponding ground-true/reconstructed images by SpeckleNet are in Columns II, IV, VI, and VIII. Presentations in Row (a) are the $d_0$-sampled speckle patterns; Row (b) the $d_5$-sampled speckle patterns and the Nearest interpolation; Row (c) the Bilinear interpolation; Row (d) the Bicubic interpolation; Row (e) the InterNet(com)-1 interpolation; Row (f) the InterNet(cc)-1 interpolation; Row (g) the InterNet(com)-2 interpolation; Row (h) the InterNet(cc)-2 interpolation. Inset quantities are formatted as PCC (MSE). PCC: Pearson's correlation coefficient between the interpolated output and target (b); MSE: mean square error between the interpolated output and target (b).

## 7. Quantitative Summary

Quantitative analysis is also performed as shown in Fig. 8. Before interpolation, the $C_m$ generally increases with the sampling pitch (Fig. 8a), since the speckle pattern nuances are diminished and become more and more alike with each other due to pixel binning. Depending on the influence to the $C_m$, the interpolation methods employed in this study can be divided into two categories: for classic methods, all realizations enlarge the $C_m$ (above the solid line) from the down-sampled speckle patterns; for learning-based methods, the mutual correlations are below the solid line. Red symbols are cases that lead to successful image reconstruction with the referred interpolation method, tending to have low $C_m$ approaching to the level with Nyquist criterion. In this scenario, the interpolated patterns within each training set (the nominal sample number is 19,800) are sufficiently different from each, thus the features for the $d_0$-trained SpeckleNet can be well identified, which permits clear imaging reconstruction. On the other hand, cases with high $C_m$ suggest the interpolated patterns within each training set are more akin, significantly reducing the number of independent samples (similar patterns are essentially one pattern from the perspective of networking training). In this scenario, the features for the $d_0$-trained SpeckleNet may be insufficient and fail to predict the hidden object information, while study for further sophisticated training is beyond this manuscript.

The other two parameters, PCC (Fig. 8b) and MSE (Fig. 8c), depict the issue from another perspective. High PCC and low MSE both suggest good match between the interpolated patterns with the original pattern that is well sampled at $d_0$ (such as Fig. 4a). As seen, the classic interpolation methods lead to low PCC values (<0.4), failing to retrieve object information by feeding their interpolated patterns into the SpeckleNet. With our learning-based interpolation methods, the PCC values are significantly higher (>0.7). This alone, however, cannot ensure successful imaging reconstruction; low MSE value (e.g., <0.01) is another criterion that must be met. Also, to be noted that for very sparsely sampled patterns (with sampling pitches of $d_6$ and $d_7$ in this study), even though both PCC and MSE of the interpolated patterns can be favorable, the trained SpeckleNet fails. This is probably because too much object information has been filtered by pixel binning during the down-sampling process, which may already have been beyond the capability of the network. Therefore, the structure and effectiveness of the proposed interpolation networks are discussed next.

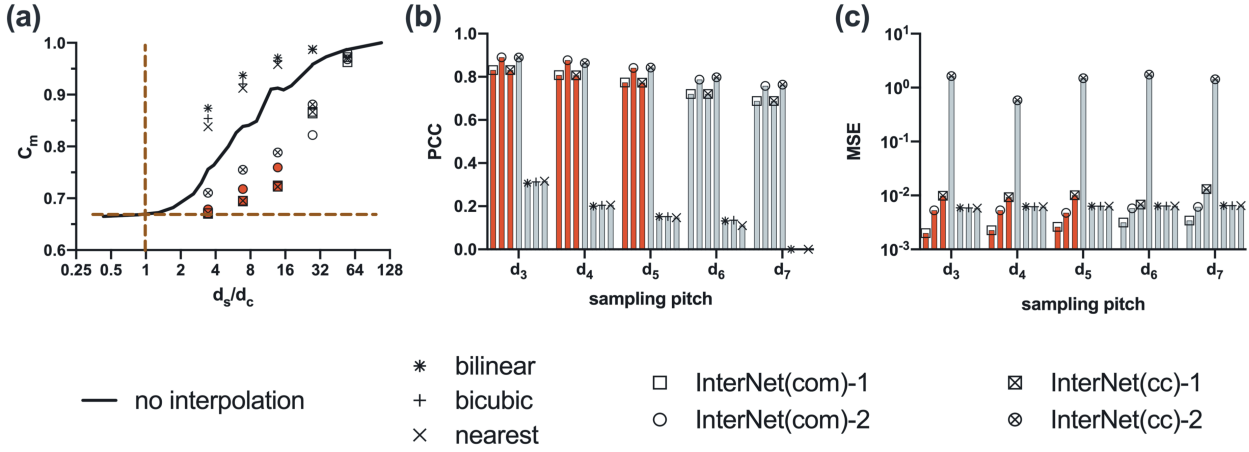

Fig. 8 (a) Mutual correlation ($C_m$) of the speckle patterns in the test dataset before and after interpolation: the solid line represents the speckle patterns before interpolation and the scattered points after interpolation. (b) and (c) are the evaluations of PCC and MSE between the interpolated patterns and the target ($d_0$-sampled, Fig. 4a). The red symbols and points in (a) and bars in (b)-(c) represent the cases that lead to successful image reconstruction with the referred interpolation method.

## 8. Interpolation network analysis

The architecture of DNN plays a role on interpolation. InterNet-1 and InterNet-2 are both based on DeeplabV3+ with a ResNet105 as backbone [36] (Fig. 2). Type-1 (InterNet-1) is based on a standard DeeplabV3+, its upsampling operation in decoder is 4x Bilinear interpolation. For Type-2 (InterNet-2), the 4x Bilinear is replaced by a densely connected transposed convolutional layer (denselayer). The denselayer has been integrated in U-net reported by Li. et al [24] to enhance the DNN generalization in image reconstruction (from speckle to object). The interpolation (from down-sampled speckle pattern to well-sampling speckle pattern), however, do not efficiently benefit from the denselayer. For performance comparison between DeeplabV3+ and U-net, readers can refer to Section 3 of Supplementary. Basically, it shows U-net is significantly less effective for speckle interpolation so that DeeplabV3+ is chosen, based on which InterNet-1 and InterNet-2 are developed in this study. When only NPCC is used as the training loss function, as shown in Fig. 6, InterNet(cc)-2 seems to results in slightly better interpolated speckle patterns, for example with ~10% higher PCC (Fig. 6VIII) than that with InterNet-1 (Fig. 6IV). However, the interpolation MSE with InterNet(cc)-2 is nearly two magnitudes larger than that with its peer (Fig. 8c). It turns out that the denselayer built in the network may constraint the short-range correlation well for the task of interpolation, but not to the long-range correlation as it concatenates more low-level features [39]. As a result, InterNet(cc)-1 can produces reconstructable interpolated patterns with minor defects (Fig. 7f), but InterNet(cc)-2 fails (Fig. 7h) as the object information has been smeared out by long-range correlation during the networking processing. Significance of selecting an appropriate architecture is undisputed [40].

The choice of training loss function is another determinant. With comloss as the training loss function, the network output is aimed to converge to minimum NPCC and MSE, which is equivalent to maximizing the PCC. The interpolated patterns generated by a well-trained InterNet(com) can be evaluated with high PCC and small MSE (Fig. 6II & VI), where the constraint to long-range correlation is applied and the object information can now be retrieved, even with the presence of denselayer in InterNet(com)-2 (Fig. 6VII and Fig. 7g). With NPCC alone as the loss function, Type-2 interpolation (without constraints to long-range correlation) from speckles of the same sampling pitch generates similar speckle morphology recovery and PCC but different orders of magnitudes of MSE (Fig. 6VI vs. VIII), leading to drastically different performances in imaging reconstruction. The influence of training loss function can also be seen from the same type of interpolations, for example InterNet(com)-1 and InterNet(cc)-1. Both lead to successful speckle pattern morphology recovery and object information retrieval (Fig. 7e-f) with similar PCCs, but with the same dataset the MSE is apparently higher and the reconstructed image is a bit noisier when NPCC alone is used as the loss function.

In brief summary, a high PCC assists to confine the short-range correlation (related to the overall morphology) of speckle patterns, but NPCC alone is not sufficient to determine an exclusive speckle pattern and ensure effective interpolation. Additional MSE in the training loss function helps to constrain the long-range correlation (related to the deterministic spatial distribution of speckle grains) of speckle patterns by suppressing the pixel-to-pixel difference between the interpolated patterns and the target during the training process. But small MSE alone is not desired, either, as can be seen from the results with classic methods (Fig. 5 and Fig. 7b-d): the MSEs are sufficiently low, but none has achieved successful speckle pattern morphology recovery or object reconstruction. Therefore, a combination of NPCC and MSE is preferred as the training loss function in this study.

## 9. Conclusion

In this work, for sub-Nyquist sampled speckle patterns, we propose InterNet, a DNN-based network, to effectively interpolate them to well-resolved speckle patterns for the first time, which is impossible with classic interpolation methods. With supervised learning, the down sampling is reversed and the loss of information due to sparse sampling can be effectively retrieved, which is validated by SpeckleNet via image reconstruction. Typical analytical solutions or non-learning-based methods are unsupervised, without a target to guide the interpolation; neither the morphology nor information can be recovered. In comparison, the supervised learning can partially overcome the current limitation of theoretical framework.

The performance of interpolation network has been significantly improved based on a standard DeeplabV3+ with a combination training loss function including NPCC and MSE, effectively constraining and converging the interpolated output with information retraced. As a result, the proposed learning-based networks provide a robust and promising strategy to retrieve optical information from down-sampled speckle patterns. The lowest sampling frequency demonstrated in this study is more than one order (~14) below the Nyquist criterion, and the corresponding speckle patterns can be interpolated by a factor of 32 with recovered object information (equivalently, resolution improvement of 32 times).

Such a capability can significantly boost the speed of data acquisition, storage, transfer, and processing of massive speckled optical signals. Notably, the application of the DNN based interpolation can be generalized due to its independence from the image reconstruction procedure. In practice, target objects to be imaged or sensed may be unknown, which impedes regular DNN training for image reconstruction. The proposed interpolation, however, exploits speckle patterns only. Therefore, the training for InterNet can be proceeded even in the absence of the knowledge of the target objects. Such a feature can significantly benefit many biomedical application scenarios, where the aim is to reveal inhomogeneities that are physically inaccessible. Last but not the least, the study successfully breaks the inherent trade-off between the FOV and resolution in speckled image reconstruction, opening up new avenues for seeing big and seeing clearly simultaneously in complex scenarios.

**Funding**. the Hong Kong Research Grant Council (No. 25204416), the Hong Kong Innovation and Technology Commission (no. ITS/022/18), the National Natural Science Foundation of China (No. 81671726, No. 81627805, and No. 81930048), Guangdong Science and Technology Commission (2019A1515011374), the Shenzhen Science and Technology Innovation Commission (No. JCYJ20170818104421564), and the A*STAR SERC AME Program: Nanoantenna Spatial Light Modulators for Next Generation Display Technologies (Grant No. A18A7b0058).

**Conflict of interest**. The authors declare no conflict of interest.

**Authors contributions.** H. L., Z. Y, and Y. L. worked on simulation. H. L. and Z. Y. contributed to the experiments. L. V. W., Y. Z., and P. L. conceived and supervised the research. All authors contributed to results discussion, manuscript writing and revision.

See Supplement 1 for supporting content.

# Reference

1. J. W. Goodman, *Speckle phenomena in optics: theory and applications* (Roberts and Company Publishers, 2007).
2. E. Akkermans and G. Montambaux, *Mesoscopic physics of electrons and photons* (Cambridge University Press, Cambridge, 2007), pp. xviii, 588 p.
3. O. Katz, P. Heidmann, M. Fink, and S. Gigan, "Non-invasive single-shot imaging through scattering layers and around corners via speckle correlations," Nat. Photon. **8**, 784-790 (2014).
4. T. Wu, O. Katz, X. Shao, and S. Gigan, "Single-shot diffraction-limited imaging through scattering layers via bispectrum analysis," Opt Lett **41**, 5003-5006 (2016).
5. J. Bertolotti, E. G. van Putten, C. Blum, A. Lagendijk, W. L. Vos, and A. P. Mosk, "Non-invasive imaging through opaque scattering layers," Nature **491**, 232-234 (2012).
6. I. M. Vellekoop and A. Mosk, "Focusing coherent light through opaque strongly scattering media," Opt. Lett. **32**, 2309-2311 (2007).
7. P. Lai, L. Wang, J. W. Tay, and L. V. Wang, "Photoacoustically guided wavefront shaping for enhanced optical focusing in scattering media," Nat. Photon. **9**, 126-132 (2015).
8. J. Xu, H. W. Ruan, Y. Liu, H. J. Zhou, and C. H. Yang, "Focusing light through scattering media by transmission matrix inversion," Opt. Express **25**, 27234-27246 (2017).
9. S. Popoff, G. Lerosey, R. Carminati, M. Fink, A. Boccara, and S. Gigan, "Measuring the transmission matrix in optics: an approach to the study and control of light propagation in disordered media," Phys. Rev. Lett. **104**, 100601 (2010).
10. H. Yu, K. Lee, and Y. Park, "Ultrahigh enhancement of light focusing through disordered media controlled by mega-pixel modes," Opt. Express **25**, 8036-8047 (2017).
11. Y. Liu, P. Lai, C. Ma, X. Xu, A. A. Grabar, and L. V. Wang, "Optical focusing deep inside dynamic scattering media with near-infrared time-reversed ultrasonically encoded (TRUE) light," Nat. Commun. **6**, 5904 (2015).
12. Z. Yu, J. Huangfu, F. Zhao, M. Xia, X. Wu, X. Niu, D. Li, P. Lai, and D. Wang, "Time-reversed magnetically controlled perturbation (TRMCP) optical focusing inside scattering media," Sci Rep **8**, 2927 (2018).
13. Z. Yu, M. Xia, H. Li, T. Zhong, F. Zhao, H. Deng, Z. Li, D. Li, D. Wang, and P. Lai, "Implementation of digital optical phase conjugation with embedded calibration and phase rectification," Sci Rep **9**, 1537 (2019).
14. Z. Yaqoob, D. Psaltis, M. S. Feld, and C. Yang, "Optical phase conjugation for turbidity suppression in biological samples," Nat. Photon. **2**, 110-115 (2008).
15. K. Si, R. Fiolka, and M. Cui, "Fluorescence imaging beyond the ballistic regime by ultrasound pulse guided digital phase conjugation," Nat Photonics **6**, 657-661 (2012).
16. C. Ma, X. Xu, Y. Liu, and L. V. Wang, "Time-reversed adapted-perturbation (TRAP) optical focusing onto dynamic objects inside scattering media," Nat. Photon. **8**, 931-936 (2014).
17. H. Yilmaz, E. G. van Putten, J. Bertolotti, A. Lagendijk, W. L. Vos, and A. P. Mosk, "Speckle correlation resolution enhancement of wide-field fluorescence imaging," Optica **2**, 424-429 (2015).
18. S. M. Popoff, A. Aubry, G. Lerosey, M. Fink, A. C. Boccara, and S. Gigan, "Exploiting the Time-Reversal Operator for Adaptive Optics, Selective Focusing, and Scattering Pattern Analysis," Phys. Rev. Lett. **107**, 263901 (2011).
19. T. Chaigne, J. Gateau, O. Katz, C. Boccara, S. Gigan, and E. Bossy, "Improving photoacoustic-guided optical focusing in scattering media by spectrally filtered detection," Opt. Lett. **39**, 6054-6057 (2014).
20. S. Cheng, H. Li, Y. Luo, Y. Zheng, and P. Lai, "Artificial intelligence-assisted light control and computational imaging through scattering media," Journal of Innovative Optical Health Sciences **12**(2019).
21. I. Goodfellow, Y. Bengio, and A. Courville, *Deep learning* (MIT press, 2016).
22. Y. LeCun, Y. Bengio, and G. Hinton, "Deep learning," Nature **521**, 436-444 (2015).
23. A. Sinha, J. Lee, S. Li, and G. Barbastathis, "Lensless computational imaging through deep learning," Optica **4**, 1117-1125 (2017).
24. S. Li, M. Deng, J. Lee, A. Sinha, and G. Barbastathis, "Imaging through glass diffusers using densely connected convolutional networks," Optica **5**(2018).
25. N. Borhani, E. Kakkava, C. Moser, and D. Psaltis, "Learning to see through multimode fibers," Optica **5**(2018).
26. B. Rahmani, D. Loterie, G. Konstantinou, D. Psaltis, and C. Moser, "Multimode optical fiber transmission with a deep learning network," Light Sci Appl **7**, 69 (2018).
27. Y. Li, Y. Xue, and L. Tian, "Deep speckle correlation: a deep learning approach toward scalable imaging through scattering media," Optica **5**(2018).
28. M. Lyu, H. Wang, G. Li, S. Zheng, and G. Situ, "Learning-based lensless imaging through optically thick scattering media," Advanced Photonics **1**(2019).
29. Y. Shen, Y. Liu, C. Ma, and L. V. Wang, "Sub-Nyquist sampling boosts targeted light transport through opaque scattering media," Optica **4**, 97-102 (2017).
30. H. Wang, Y. Rivenson, Y. Jin, Z. Wei, R. Gao, H. Gunaydin, L. A. Bentolila, C. Kural, and A. Ozcan, "Deep learning enables cross-modality super-resolution in fluorescence microscopy," Nat Methods **16**, 103-110 (2019).
31. Y. Rivenson, Z. Göröcs, H. Günaydin, Y. Zhang, H. Wang, and A. Ozcan, "Deep learning microscopy," Optica **4**(2017).

32. S. Popoff, G. Lerosey, M. Fink, A. C. Boccara, and S. Gigan, "Image transmission through an opaque material," Nat. Commun. **1**, 81 (2010).
33. W. Xiong, C. W. Hsu, and H. Cao, "Long-range spatio-temporal correlations in multimode fibers for pulse delivery," Nat Commun **10**, 2973 (2019).
34. J. W. Goodman, *Statistical optics* (John Wiley & Sons, 2015).
35. J. Pawley, *Handbook of biological confocal microscopy* (Springer Science & Business Media, 2010).
36. L.-C. Chen, G. Papandreou, F. Schroff, and H. Adam, "Rethinking atrous convolution for semantic image segmentation," arXiv preprint arXiv:1706.05587 (2017).
37. I. Starshynov, A. M. Paniagua-Diaz, N. Fayard, A. Goetschy, R. Pierrat, R. Carminati, and J. Bertolotti, "Non-Gaussian Correlations between Reflected and Transmitted Intensity Patterns Emerging from Opaque Disordered Media," Physical Review X **8**(2018).
38. G. Barbastathis, A. Ozcan, and G. Situ, "On the use of deep learning for computational imaging," Optica **6**(2019).
39. Z. L. Gao Huang, Laurens van der Maaten and Kilian Q. Weinberger, "Densely Connected Convolutional Networks," (2016).
40. A. M. Zador, "A critique of pure learning and what artificial neural networks can learn from animal brains," Nat Commun **10**, 3770 (2019).